\documentclass[longauth, letter, traditabstract]{aa} 
\usepackage{graphicx}
\usepackage{txfonts}
\usepackage{fixltx2e}
\usepackage{amssymb}
\usepackage{natbib}
\usepackage{rotating}
\bibpunct{(}{)}{;}{a}{}{,}

\begin{document}
\title{\textit{Herschel}-HIFI observations of high-\textit{J} CO lines in the NGC 1333 low-mass star-forming region\thanks{Herschel is an ESA space observatory with science instruments provided by European-led Principal Investigator consortia and with important participation from NASA.}}
  \titlerunning{HIFI observations of high-$J$ CO in NGC~1333}

\author{
     U.A.~Y{\i}ld{\i}z\inst{\ref{inst1}}
\and E.F.~van~Dishoeck\inst{\ref{inst1},\ref{inst2}}
\and L.E.~Kristensen\inst{\ref{inst1}}
\and R.~Visser\inst{\ref{inst1}}
\and J.K.~J{\o}rgensen\inst{\ref{inst3}}
\and G.J.~Herczeg\inst{\ref{inst2}}
\and T.A.~van~Kempen\inst{\ref{inst1},\ref{inst4}}
\and M.R.~Hogerheijde\inst{\ref{inst1}}
\and S.D.~Doty\inst{\ref{inst5}}
\and A.O.~Benz\inst{\ref{inst6}}
\and S.~Bruderer\inst{\ref{inst6}}
\and S.F.~Wampfler\inst{\ref{inst6}}
\and E.~Deul\inst{\ref{inst1}}
\and R.~Bachiller\inst{\ref{inst7}}
\and A.~Baudry\inst{\ref{inst8}}
\and M.~Benedettini\inst{\ref{inst9}}
\and E.~Bergin\inst{\ref{inst10}}
\and P.~Bjerkeli\inst{\ref{inst11}}
\and G.A.~Blake\inst{\ref{inst12}}
\and S.~Bontemps\inst{\ref{inst8}}
\and J.~Braine\inst{\ref{inst8}}
\and P.~Caselli\inst{\ref{inst13},\ref{inst14}}
\and J.~Cernicharo\inst{\ref{inst15}}
\and C.~Codella\inst{\ref{inst14}}
\and F.~Daniel\inst{\ref{inst15}}
\and A.M.~di~Giorgio\inst{\ref{inst9}}
\and C.~Dominik\inst{\ref{inst16},\ref{inst17}}
\and P.~Encrenaz\inst{\ref{inst18}}
\and M.~Fich\inst{\ref{inst19}}
\and A.~Fuente\inst{\ref{inst20}}
\and T.~Giannini\inst{\ref{inst21}}
\and J.R.~Goicoechea\inst{\ref{inst15}}
\and Th.~de~Graauw\inst{\ref{inst22}}
\and F.~Helmich\inst{\ref{inst22}}
\and F.~Herpin\inst{\ref{inst8}}
\and T.~Jacq\inst{\ref{inst8}}
\and D.~Johnstone\inst{\ref{inst23},\ref{inst24}}
\and B.~Larsson\inst{\ref{inst25}}
\and D.~Lis\inst{\ref{inst26}}
\and R.~Liseau\inst{\ref{inst11}}
\and F.-C.~Liu\inst{\ref{inst27}}
\and M.~Marseille\inst{\ref{inst22}}
\and C.~M$^{\textrm c}$Coey\inst{\ref{inst19},\ref{inst28}}
\and G.~Melnick\inst{\ref{inst4}}
\and D.~Neufeld\inst{\ref{inst29}}
\and B.~Nisini\inst{\ref{inst21}}
\and M.~Olberg\inst{\ref{inst11}}
\and B.~Parise\inst{\ref{inst27}}
\and J.C.~Pearson\inst{\ref{inst30}}
\and R.~Plume\inst{\ref{inst31}}
\and C.~Risacher\inst{\ref{inst22}}
\and J.~Santiago-Garc\'{i}a\inst{\ref{inst32}}
\and P.~Saraceno\inst{\ref{inst9}}
\and R.~Shipman\inst{\ref{inst22}}
\and M.~Tafalla\inst{\ref{inst7}}
\and A.~G.~G.~M.~Tielens\inst{\ref{inst1}}
\and F.~van der Tak\inst{\ref{inst22},\ref{inst33}}
\and F.~Wyrowski\inst{\ref{inst27}}
\and P.~Dieleman\inst{\ref{inst22}}
\and W.~Jellema\inst{\ref{inst22}}
\and V.~Ossenkopf\inst{\ref{inst34}}
\and R.~Schieder\inst{\ref{inst34}}
\and J.~Stutzki\inst{\ref{inst34}}
}

\institute{
Leiden Observatory, Leiden University, PO Box 9513, 2300 RA Leiden, The Netherlands, \label{inst1} \email{yildiz@strw.leidenuniv.nl}
\and
Max Planck Institut f\"{u}r Extraterrestrische Physik, Giessenbachstrasse 1, 85748 Garching, Germany\label{inst2}
\and
Centre for Star and Planet Formation, Natural History Museum of Denmark, University of Copenhagen,
{\O}ster Voldgade 5-7, DK-1350 Copenhagen K., Denmark\label{inst3}
\and
Harvard-Smithsonian Center for Astrophysics, 60 Garden Street, MS 42, Cambridge, MA 02138, USA\label{inst4}
\and
Department of Physics and Astronomy, Denison University, Granville, OH, 43023, USA\label{inst5}
\and
Institute of Astronomy, ETH Zurich, 8093 Zurich, Switzerland\label{inst6}
\and
Observatorio Astron\'{o}mico Nacional (IGN), Calle Alfonso XII,3. 28014 Madrid, Spain\label{inst7}
\and
Universit\'{e} de Bordeaux, Laboratoire d'Astrophysique de Bordeaux, France; CNRS/INSU, UMR 5804, Floirac, France\label{inst8}
\and
INAF - Istituto di Fisica dello Spazio Interplanetario, Area di Ricerca di Tor Vergata, via Fosso del Cavaliere 100, 00133 Roma, Italy\label{inst9}
\and
Department of Astronomy, The University of Michigan, 500 Church Street, Ann Arbor, MI 48109-1042, USA\label{inst10}
\and
Department of Radio and Space Science, Chalmers University of Technology, Onsala Space Observatory, 439 92 Onsala, Sweden\label{inst11}
\and
California Institute of Technology, Division of Geological and Planetary Sciences, MS 150-21, Pasadena, CA 91125, USA\label{inst12}
\and
School of Physics and Astronomy, University of Leeds, Leeds LS2 9JT, UK\label{inst13}
\and
INAF - Osservatorio Astrofisico di Arcetri, Largo E. Fermi 5, 50125 Firenze, Italy\label{inst14}
\and
Centro de Astrobiolog\'{\i}a. Departamento de Astrof\'{\i}sica. CSIC-INTA. Carretera de Ajalvir, Km 4, Torrej\'{o}n de Ardoz., 28850 Madrid, Spain\label{inst15}
\and
Astronomical Institute Anton Pannekoek, University of Amsterdam, Kruislaan 403, 1098 SJ Amsterdam, The Netherlands\label{inst16}
\and
Department of Astrophysics/IMAPP, Radboud University Nijmegen, P.O. Box 9010, 6500 GL Nijmegen, The Netherlands\label{inst17}
\and
LERMA and UMR 8112 du CNRS, Observatoire de Paris, 61 Av. de l'Observatoire, 75014 Paris, France\label{inst18}
\and
University of Waterloo, Department of Physics and Astronomy, Waterloo, Ontario, Canada\label{inst19}
\and
Observatorio Astron\'{o}mico Nacional, Apartado 112, 28803 Alcal\'{a} de Henares, Spain\label{inst20}
\and
INAF - Osservatorio Astronomico di Roma, 00040 Monte Porzio catone, Italy\label{inst21}
\and
SRON Netherlands Institute for Space Research, PO Box 800, 9700 AV, Groningen, The Netherlands\label{inst22}
\and
National Research Council Canada, Herzberg Institute of Astrophysics, 5071 West Saanich Road, Victoria, BC V9E 2E7, Canada\label{inst23}
\and
Department of Physics and Astronomy, University of Victoria, Victoria, BC V8P 1A1, Canada\label{inst24}
\and
Department of Astronomy, Stockholm University, AlbaNova, 106 91 Stockholm, Sweden\label{inst25}
\and
California Institute of Technology, Cahill Center for Astronomy and Astrophysics, MS 301-17, Pasadena, CA 91125, USA\label{inst26}
\and
Max-Planck-Institut f\"{u}r Radioastronomie, Auf dem H\"{u}gel 69, 53121 Bonn, Germany\label{inst27}
\and
the University of Western Ontario, Department of Physics and Astronomy, London, Ontario, N6A 3K7, Canada\label{inst28}
\and
Department of Physics and Astronomy, Johns Hopkins University, 3400 North Charles Street, Baltimore, MD 21218, USA\label{inst29}
\and
Jet Propulsion Laboratory, California Institute of Technology, Pasadena, CA 91109, USA\label{inst30}
\and
Department of Physics and Astronomy, University of Calgary, Calgary, T2N 1N4, AB, Canada\label{inst31}
\and
Instituto de Radioastronom\'{i}a Milim\'{e}trica (IRAM), Avenida Divina Pastora 7, N\'{u}cleo Central, E-18012 Granada, Spain\label{inst32}
\and
Kapteyn Astronomical Institute, University of Groningen, PO Box 800, 9700 AV, Groningen, The Netherlands\label{inst33}
\and
KOSMA, I. Physik. Institut, Universit\"{a}t zu K\"{o}ln, Zülpicher Str. 77, D 50937 K\"{o}ln, Germany\label{inst34}
}

   \date{Accepted: 2010 Aug 2}

\def\placeFigureIRASSpectraOnTop{
\begin{figure}
\begin{center}
\includegraphics[scale=0.36, angle=270]{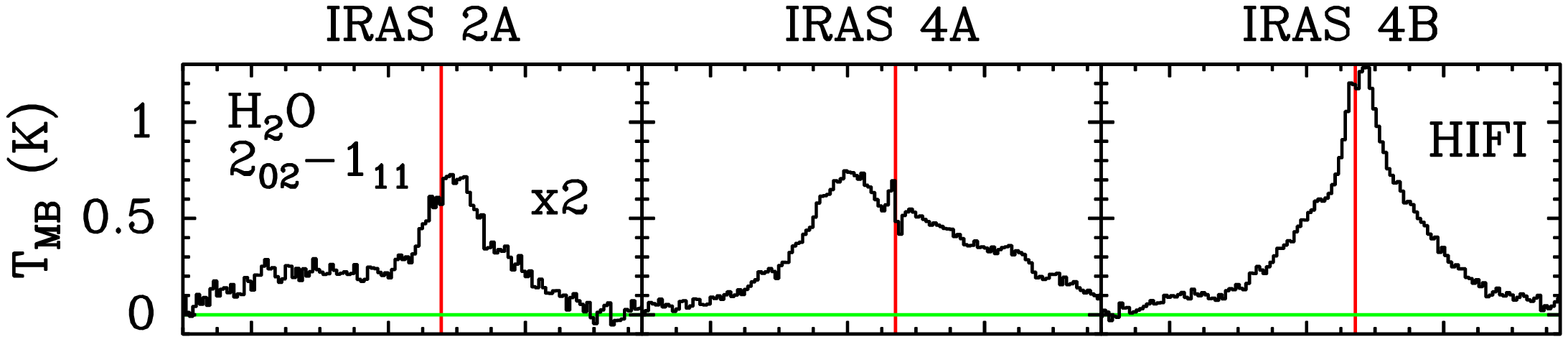}

\smallskip
\includegraphics[scale=0.41, angle=270]{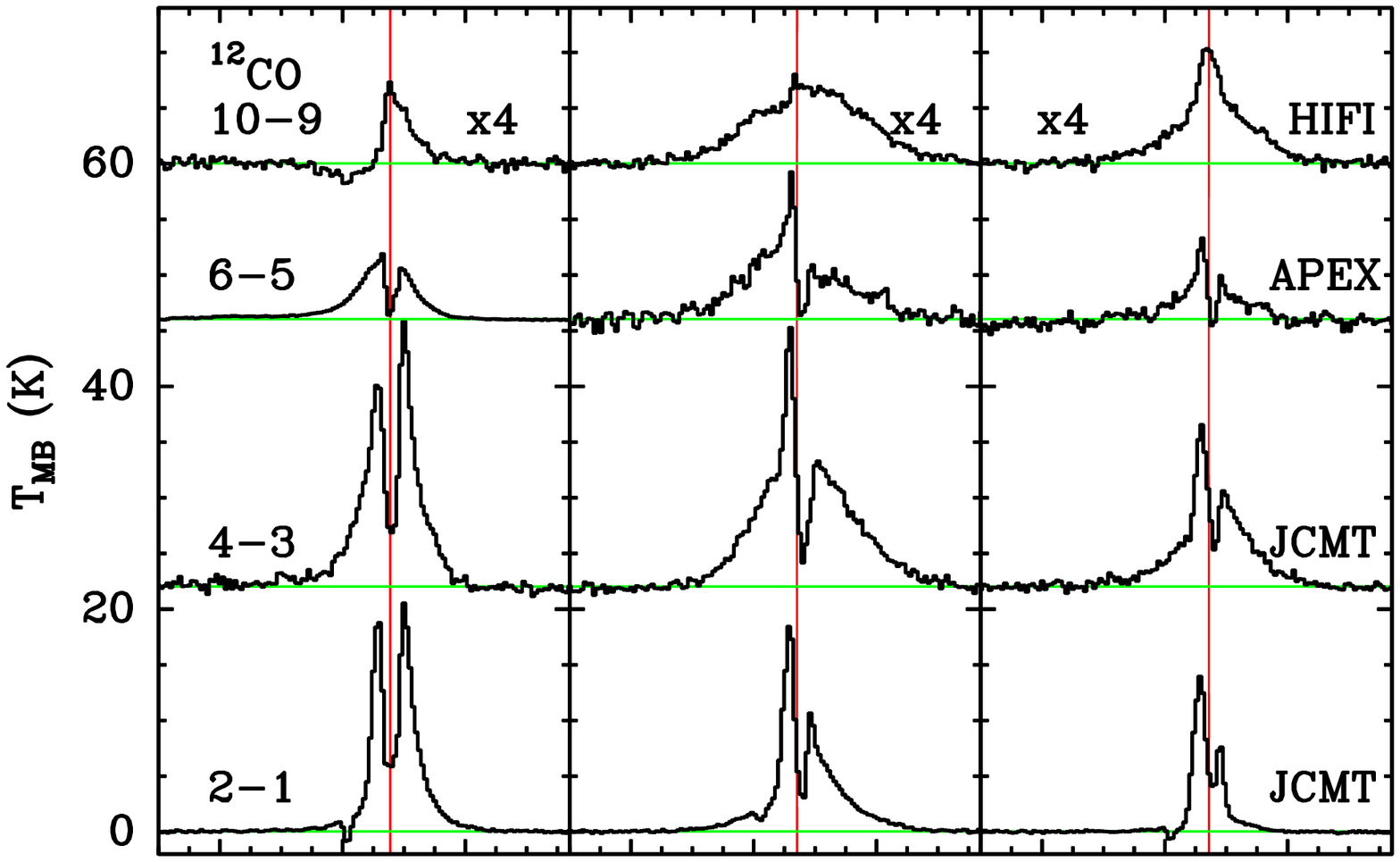}

\smallskip
\includegraphics[scale=0.36, angle=270]{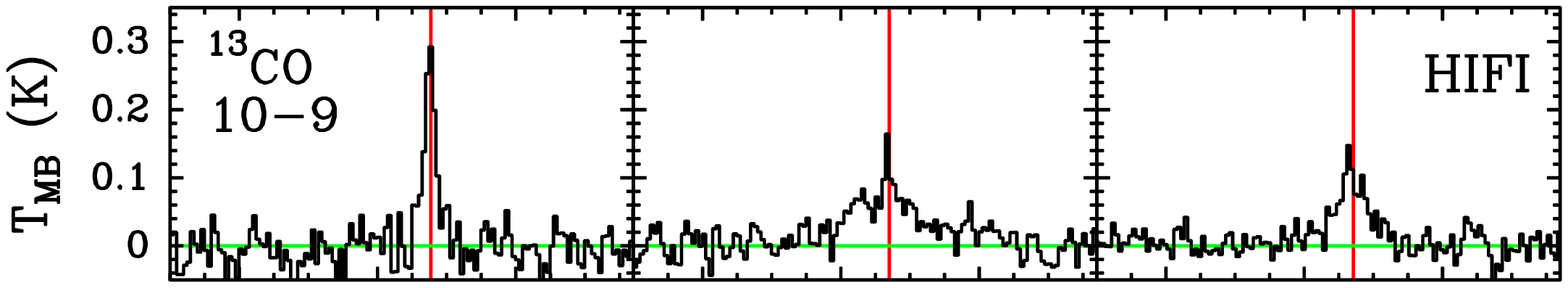}

\smallskip
\includegraphics[scale=0.36, angle=270]{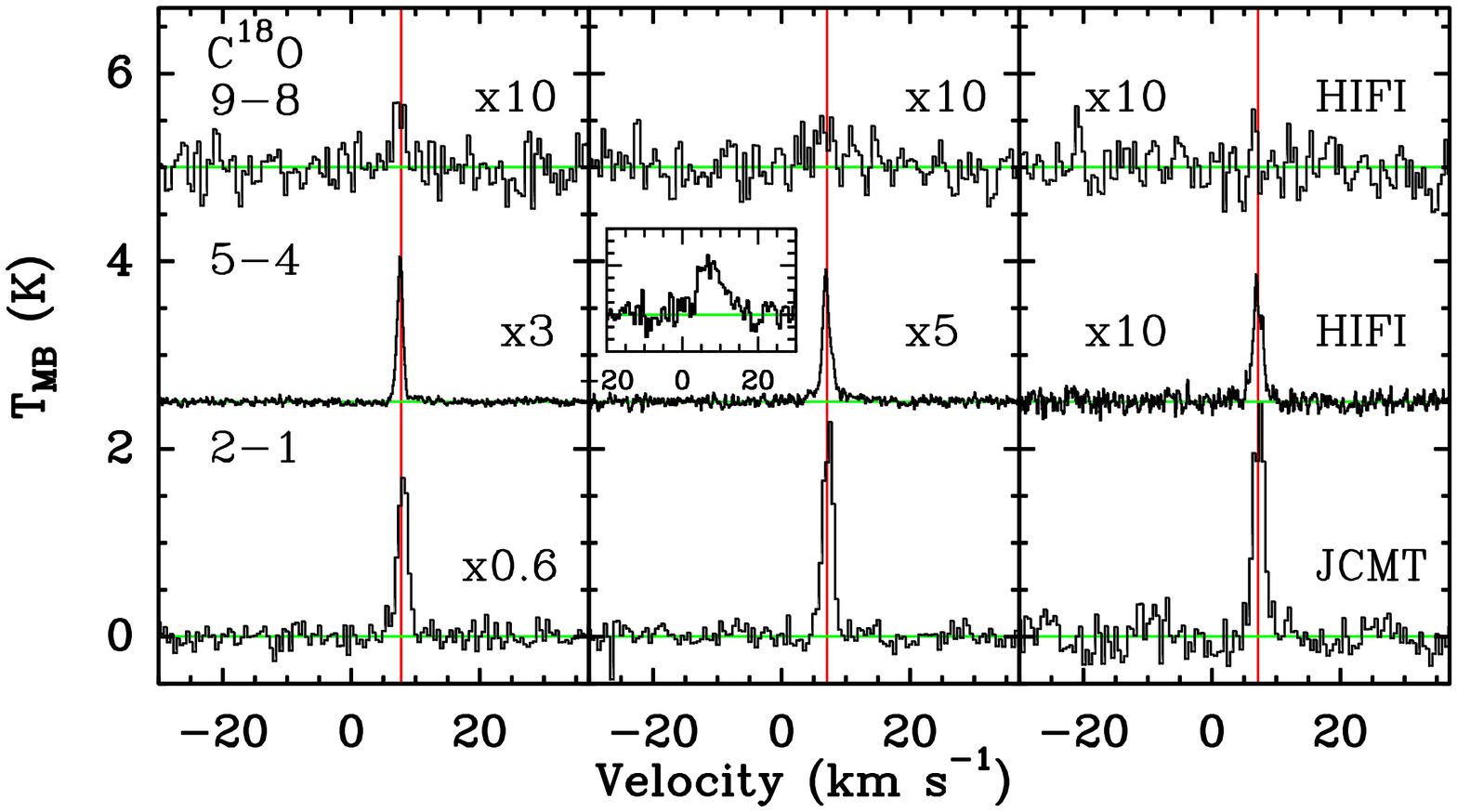}

\end{center}
\caption{Spectra at the central positions of  IRAS~2A, 4A and 4B. \textit{Top to bottom:} \mbox{H$_{2}$O~2$_{02}$--1$_{11}$} line from \citet{Kristensen10} illustrating the medium and broad components, and spectra of \mbox{\element[][12]{C}O}, \mbox{\element[][13]{CO}}, and \mbox{C\element[][18]{O}}. The red lines correspond to the source velocities as obtained from the low-$J$ \mbox{C\element[][18]{O}} lines. The insert in the C$^{18}$O 5--4 line for IRAS~4A illustrates the weak medium component with peak $T_{\rm MB}=22$ mK obtained after subtracting a Gaussian fit to the narrow line.}
\label{3:fig:AllSpectraOntop}
\end{figure}
}

\def\placeFigureRatios{
\begin{figure}
\begin{center}
\includegraphics[scale=0.35, angle=270]{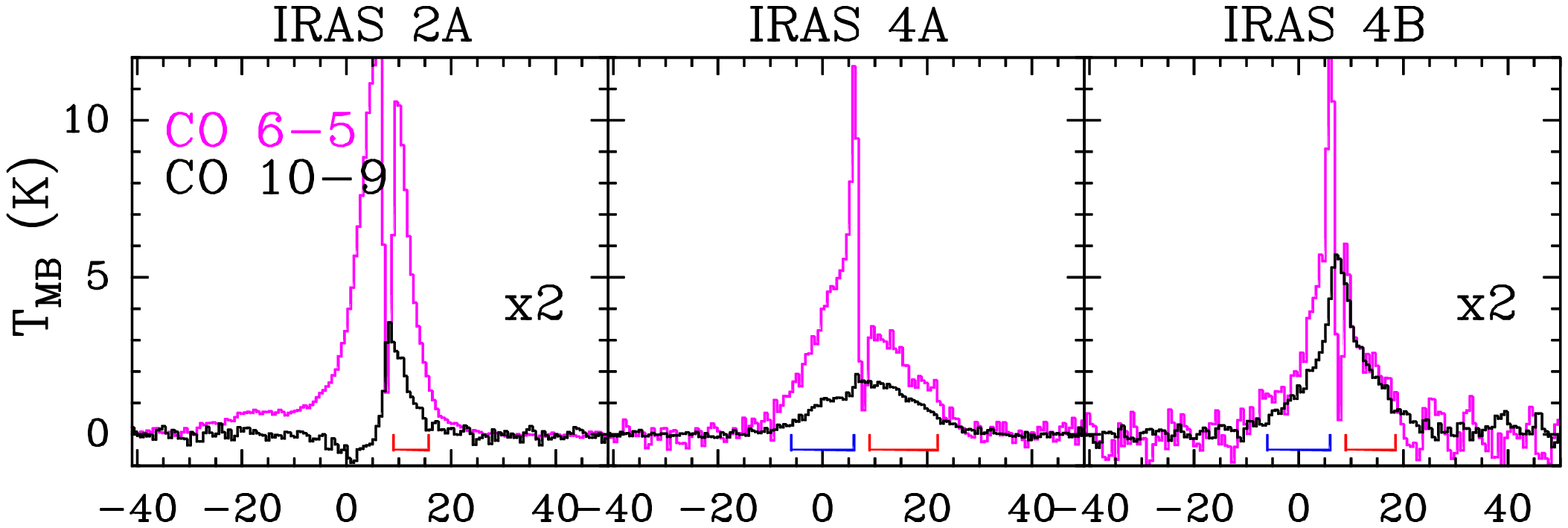}
\includegraphics[scale=0.35, angle=270]{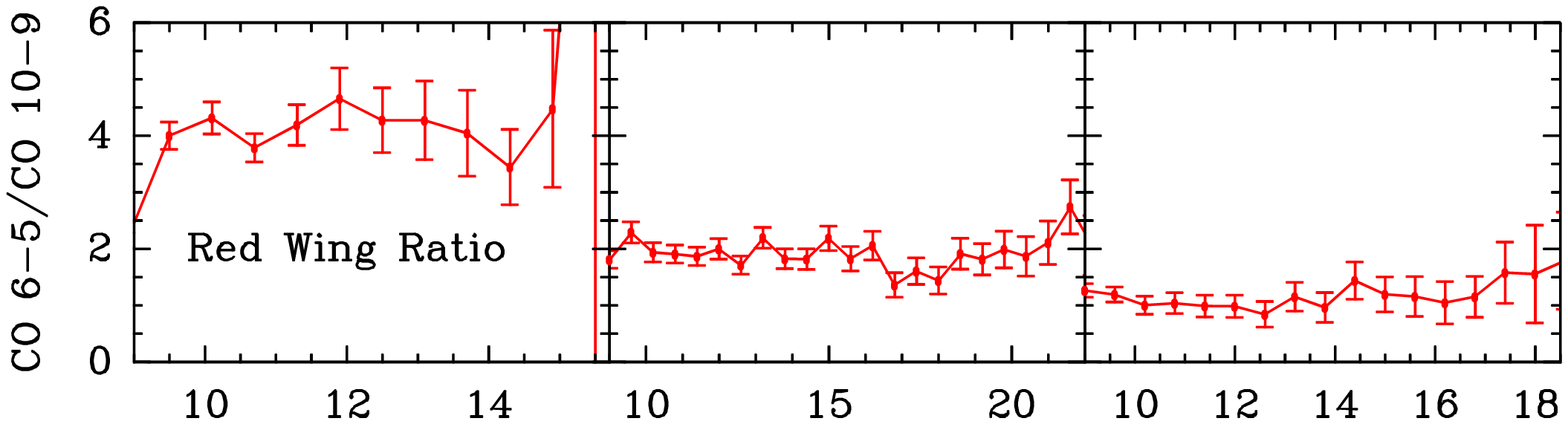}
\includegraphics[scale=0.35, angle=270]{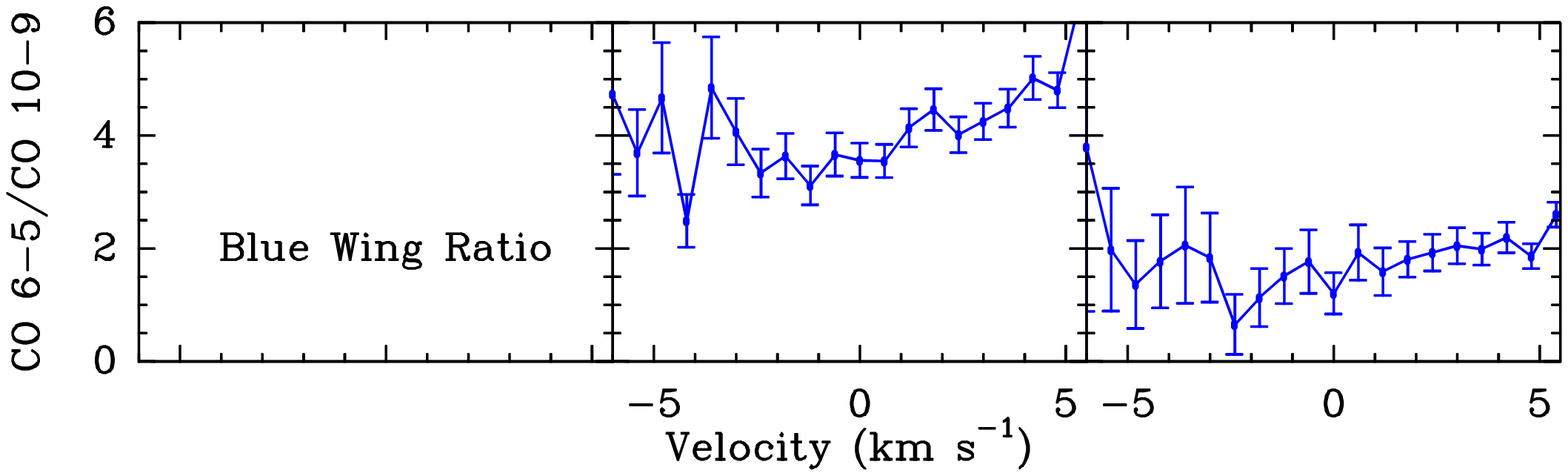}
\end{center}
\caption{Ratios of \mbox{\element[]{C}O 6--5}/\mbox{\element[]{CO} 10--9}. \textit{Top}: CO line profiles. 
The \mbox{CO~6--5} and 10--9 profiles have been multiplied
by a factor of 2 for IRAS 2A and 4B. 
\textit{Middle} and \textit{bottom}: ratio of line wing intensity in the specified 
velocity range indicated in the top panel for the red and blue wings.}
\label{3:fig:ratios}
\end{figure}
}

\def\placeFigureRatran{
\begin{figure}
\begin{center}
\includegraphics[scale=0.40]{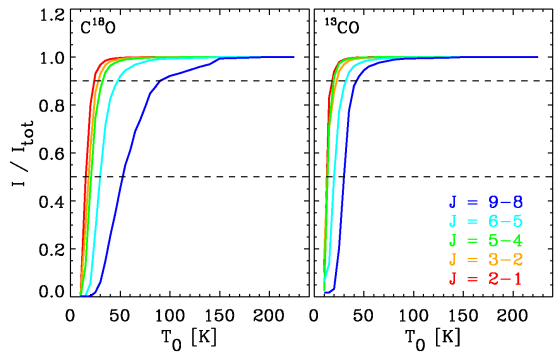}
\end{center}
\caption{Dependence of line intensities on  temperature
  $T_0$ of C$^{18}$O (left) and $^{13}$CO (right) for an ``anti-jump"
  model of the CO abundance in the IRAS~2A envelope. The line
  intensities are measured relative to a model where the CO abundance
  is undepleted at all radii. Each curve therefore represents the
  fraction of the line intensity for the given transition, which has
  its origin in gas at temperatures below $T_0$. The dashed lines
  indicate the levels corresponding to 50 and 90$\%$ respectively.}
\label{3:fig:RatranModel}
\end{figure}
}



\def\placeTableOverviewOfTheObservations{
\begin{table}
\caption{Overview of the observations of IRAS~2A, 4A, and 4B.}
\tiny
\begin{center}
\begin{tabular}{l r r r l r l}
\hline \hline
Mol. & Trans.& $E_\mathrm{u}/k_{\mathrm{B}}$ & Frequency & Tel./Inst. & Beam  &  Ref. \\ 
 &  & (K) & (GHz)  & & size ($\arcsec$) & \\ 
\hline
 CO                         & 2--1   & 16.6    & 230.538   & JCMT      & 22   & 1\\
                               & 4--3   & 55.3    & 461.041   & JCMT      & 11   & 2\\
                               & 6--5   & 116.2  & 691.473   & APEX      & 9     & 3\\
                               & 10--9 & 304.2  & 1151.985 & HIFI-5a  & 20   & 4\\
 \element[][13]{C}O & 10--9 & 290.8  & 1101.349 & HIFI-4b  & 21   & 4\\
 C\element[][18]{O} & 1--0   &  5.3     & 109.782   & Onsala   & 34   & 1\\
                               & 2--1   & 15.8    & 219.560   & JCMT      & 23   & 1\\
                               & 3--2   & 31.6    & 329.331   & JCMT      & 15   & 1\\
                               & 5--4   & 79.0    & 548.831   & HIFI-1a  & 42   & 4\\
                               & 6--5   & 110.6  & 658.553   & APEX      & 10   & 3\\
                               & 9--8   & 237.0  & 987.560   & HIFI-4a  & 23   & 4\\
                               &10--9  & 289.7  &1097.162  & HIFI-4b  & 21   & 4\\
\hline
\end{tabular}
\end{center}
\label{tbl:overviewobs}
(1) \citet{Jorgensen02}; (2) JCMT archive; (3) Y{\i}ld{\i}z et al. (in prep.); (4) this work
\end{table}
}

\def\placeTableColumnDensities{
\begin{table}
\caption{Summary of column densities, $N$(H$_2$) in cm$^{-2}$ in the broad and
medium components in 20$''$ beam.}
\tiny
\begin{center}
\begin{tabular}{l c c c}
\hline \hline
Source & Broad Comp. & Medium Comp. \\
\hline
IRAS~2A  &   $6\times10^{19}$ \tablefootmark{a}   &   $2\times10^{20}$  \tablefootmark{b}    \\
IRAS~4A  &    $4\times10^{20}$ \tablefootmark{b}  &   $6\times10^{20}$  \tablefootmark{c}   \\
IRAS~4B   &   $1\times10^{20}$ \tablefootmark{b}   &   $2\times10^{20}$  \tablefootmark{c} \\
\hline
\end{tabular}
\end{center}
Obtained from \tablefoottext{a}{CO 6--5}, \tablefoottext{b}{CO 10--9}, \tablefoottext{c}{$^{13}$CO 10--9 spectra.} 
\label{tbl:columndensities}
\end{table}
}

\def\placeTableObservedLineIntensities{
\begin{table}
\caption{Observed line intensities.}
\tiny
\begin{center}
\begin{tabular}{l l r r r r }
\hline \hline
Source & Mol.      & Trans. & $\int T_{\mathrm{MB}} \mathrm{d}V$ & $T_{\mathrm{peak}}$ & rms\tablefootmark{a} \\
 &  &  & (K~km~s$^{-1}$) & (K) & (K) \\
\hline
IRAS~2A & CO          & 2--1   & 127.5 & 20.5 & 0.08 \\
        &                      & 4--3   & 177.2 & 23.8 & 0.44 \\
        &                      & 6--5   & 57.0   & 5.9   & 0.11 \\
        &                      & 10--9 & 16.3   & 1.71 & 0.078 \\ 
&\element[][13]{CO} & 10--9 & 0.4     & 0.3   & 0.026\\
&C\element[][18]{O} & 1--0   & 5.6     & 4.0   & 0.27\\
        &                      & 2--1   & 5.83   & 2.3   & 0.15\\
        &                      & 3--2   & 4.7     & 3.2   & 0.13\\
        &                      & 5--4   & 0.62     & 0.46 & 0.004\\
        &                      & 6--5   & 1.8     & 1.1   & 0.11\\
        &                      & 9--8   & 0.2     & 0.07 & 0.018 \\
        &                      &10--9   & 0.15     & 0.06 & 0.017 \\
\hline
IRAS~4A & CO          & 2--1   & 117.2 & 18.4  & 0.07\\
        &                      & 4--3   & 221.1 & 23.3  & 0.32\\
        &                      & 6--5   & 121.9 & 13.2 &  0.59\\
        &                      & 10--9 & 35.7   & 1.9 & 0.073 \\
&\element[][13]{CO} & 10--9 & 1.2    & 0.2 & 0.017\\
&C\element[][18]{O} & 2--1   & 4.3    & 2.3  & 0.09\\
        &                      & 5--4   & 0.5    & 0.26 & 0.005\\
        &                      & 9--8   & 0.1    & 0.05 & 0.018 \\
\hline 
IRAS~4B & CO           & 2--1   & 54.8  & 13.9 & 0.07\\
        &                      & 4--3   & 115.2 & 14.6 & 0.26\\
        &                      & 6--5   & 43.3   & 7.3  &  0.36\\
        &                      & 10--9 & 26.8  & 2.6 & 0.076\\
&\element[][13]{CO} & 10--9 & 0.7    & 0.15 & 0.017\\
&C\element[][18]{O} & 2--1   & 4.9    & 2.5   & 0.19\\
        &                      & 5--4   & 0.23    & 0.12 & 0.005\\
        &                      & 9--8   & $<$0.07       &  - & 0.019\\
\hline
\end{tabular}
\end{center}
\tablefoottext{a}{In 0.5 km~s$^{-1}$ bins.}
\label{tbl:lineintensities}
\end{table}
}


\begin{abstract}
{\textit{Herschel}-HIFI observations of high-$J$ lines (up to $J_\mathrm{u}$=10) 
of $^{12}$CO, $^{13}$CO and C$^{18}$O are presented toward three deeply embedded 
low-mass protostars, \object{NGC~1333} \object{IRAS~2A}, \object{IRAS~4A}, and 
\object{IRAS~4B}, obtained as part of the \textit{Water In Star-forming regions 
with Herschel} (WISH) key program. The spectrally-resolved HIFI data are 
complemented by ground-based observations of lower-$J$ CO and 
isotopologue lines. The $^{12}$CO 10--9 profiles are dominated by broad 
(FWHM 25--30 km s$^{-1}$) emission. Radiative transfer models are used 
to constrain the temperature of this shocked gas to 100--200 K. Several
CO and $^{13}$CO line profiles also reveal a medium-broad component 
(FWHM 5--10 km s$^{-1}$), seen prominently in H$_2$O lines. 
Column densities for both components are presented, providing a reference 
for determining abundances of other molecules in the same gas. The narrow 
C$^{18}$O 9--8 lines probe the warmer part of the quiescent envelope. 
Their intensities require a jump in the CO abundance at an evaporation 
temperature around 25 K, thus providing new direct evidence 
for a CO ice evaporation zone around low-mass protostars.}
\end{abstract}

   \maketitle



\section{Introduction}
The earliest protostellar phase just after cloud collapse -- the so-called Class 0
phase -- is best studied at mid-infrared and longer
wavelengths \citep{Andre00}.  To understand the physical and chemical
evolution of low-mass protostars, in particular the relative
importance of radiative heating and shocks in their energy budget,
observations are required that can separate these components. The
advent of the Heterodyne Instrument for the Far-Infrared (HIFI) on
\textit{Herschel} opens up the possibility to obtain spectrally resolved
data from higher-frequency lines that are sensitive to gas
temperatures up to several hundred Kelvin.

Because of its high abundance and strong lines, CO is the primary
molecule to probe the various components of protostellar systems
(envelope, outflow, disk).  The main advantage of CO compared with
other molecules (including water) is that its chemistry is simple,
with most carbon locked up in CO in dense clouds. Also, its
evaporation temperature is low, around 20~K for pure CO ice
\citep{Collings03,Oberg05}, so that its freeze-out zone is much
smaller than that of water.  Most ground-based observations of CO and
its isotopologues have been limited to the lowest rotational
lines originating from levels up to 35 K. ISO has detected strong far-infrared CO lines up to
$J_\mathrm{u}$=29 from \mbox{Class 0} sources \citep{Giannini01} 
but the emission is spatially unresolved in the large 80$\arcsec$ beam.  ISO 
also lacked the spectral resolution needed to separate the shocked and 
quiescent gas or to detect intrinsically-weaker $^{13}$CO and C$^{18}$O 
lines on top of the strong continuum.

The NGC 1333 region in Perseus
\citep[$d=235$ pc;][]{Hirota08} contains several deeply embedded Class 0 sources
within a $\sim$1 pc region driving powerful outflows
\citep[e.g.,][]{Liseau88,Hatchell08}. 
The protostars IRAS~4A
and~4B, separated by $\sim$31$\arcsec$, and IRAS~2A are prominent
submillimeter continuum sources (luminosities of 5.8, 3.8 and \mbox{20
$L_{\odot}$}) with envelope masses of 4.5, 2.9 and 1.0 $M_{\odot}$,
respectively \citep{Sandell91,Jorgensen09}. All three are among the
brightest and best studied low-mass sources in terms of molecular
lines, with several complex molecules detected
\citep[e.g.,][]{Blake95,Bottinelli07}. Here
HIFI data of CO and its isotopologues are presented for these three sources and
used to quantify the different physical components. In an accompanying letter,
\citet{Kristensen10} present complementary HIFI observations of
H$_2$O and analyze CO/H$_2$O abundance ratios.


\section{Observations and results}
The NGC~1333 data were obtained with HIFI \citep{degraauw10} onboard the \textit{Herschel} Space Observatory
\citep{Pilbratt10}, in the context of the WISH key program
(van Dishoeck et al. in prep.). Single pointings at the three source positions
were carried out between 2010~March 3 and 15 during the
\textit{Herschel} HIFI priority
science program (PSP). Spectral lines were observed in dual-beam switch
(DBS) mode in HIFI bands 1a, 4a, 4b, and 5a with a chop reference
position located 3$\arcmin$ from the source positions. The observed positions (J2000) are: 
IRAS~2A: $3^{\mathrm{h}}28^{\mathrm{m}}55\fs6$, $+31\degr14\arcmin37\farcs1$; 
IRAS~4A: $3^{\mathrm{h}}29^{\mathrm{m}}10\fs5$, $+31\degr13\arcmin30\farcs9$; and 
IRAS~4B: $3^{\mathrm{h}}29^{\mathrm{m}}12\fs0$, $+31\degr13\arcmin08\farcs1$ \citep{Jorgensen09}.

\placeTableOverviewOfTheObservations

Table \ref{tbl:overviewobs} summarizes the lines observed with HIFI
together with complementary lower-$J$ lines obtained with
ground-based telescopes.
The \textit{Herschel} data were taken using the wide band
spectrometer (WBS) and high resolution spectrometer (HRS)
backends. Owing to the higher noise ($\sqrt{2}$ more) in HRS
than WBS, mainly WBS data are presented here. Only
the narrow \mbox{C\element[][18]{O} 5--4} lines use the HRS data.
Integration times (on+off) are 10, 20, 30, 40, and 60 minutes for the 
\mbox{\element[][12]{C}O 10--9}, \mbox{C\element[][18]{O} 9--8}, \mbox{10--9},
\mbox{\element[][13]{CO} 10--9}, and 
\mbox{C\element[][18]{O} 5--4} lines respectively. The HIFI beam sizes correspond to
$\sim$20$\arcsec$ ($\sim$4700 AU) at \mbox{1152 GHz} and $\sim$42$\arcsec$
($\sim$10000 AU) at \mbox{549 GHz}.  Except for the
\mbox{\element[][12]{C}O~10--9} line, all
isotopologue lines were observed together with H$_2$O lines.

The calibration uncertainty for the HIFI data is of the order of 20$\%$ and
the pointing accuracy is around 2$\arcsec$. The measured line
intensities were converted to the main-beam brightness temperatures
\mbox{$T_{\mathrm{MB}} = T_{A}^{*}/ \eta_{\mathrm{MB}}$} by using a
beam efficiency \mbox{$\eta_{\mathrm{MB}}=0.74$} for all HIFI lines.
Data processing started from the standard HIFI pipeline in the
\textit{Herschel} interactive processing environment
(HIPE\footnote{HIPE is a joint development by the Herschel Science
  Ground Segment Consortium, consisting of ESA, the NASA Herschel
  Science Center, and the HIFI, PACS and SPIRE consortia.}) ver.  3.0.1
\citep{OttS10}, where the $V_{\mathrm{LSR}}$ precision is 
of the order of a few m~s$^{-1}$.  
Further reduction and analysis were done using the
GILDAS-\verb1CLASS1\footnote{{http://www.iram.fr/IRAMFR/GILDAS/}}
software. The spectra from the H- and V-polarizations were averaged in
order to obtain a better $S/N$. In some cases a discrepancy of 30\%
was found between the two polarizations, in which case only the H band
spectra were used for analysis since their rms is lower.

Complementary ground-based spectral line observations of
\mbox{\element[][12]{C}O 6--5} were obtained at the 12-m Atacama
Pathfinder Experiment Telescope (APEX), using the CHAMP$^{+}$
2$\times$7 pixel array receiver \citep{2008SPIE.7020E..25G}. The
lower-$J$ spectral lines were obtained from the James Clerk Maxwell
Telescope (JCMT) archive and from
\citet{Jorgensen02}.  Details will
be presented elsewhere (Y{\i}ld{\i}z et al., in prep.).

\placeTableObservedLineIntensities

The observed line profiles are presented in Fig.
\ref{3:fig:AllSpectraOntop} and the corresponding line intensities in
Table \ref{tbl:lineintensities}. For the $^{12}$CO 10--9 toward IRAS~2A, the 
emission from the blue line wing was chopped out
due to emission at the reference position located in the blue
part of the SVS~13 outflow.
A Gaussian fitted to the
red component of the line was used to obtain the
integrated intensity.  

\placeFigureIRASSpectraOnTop

\citet{Kristensen10} identify three components in the H$_{2}$O
  line profiles centered close to the source velocities: a broad
  underlying emission profile (Gaussian with \mbox{FWHM $\sim$25--30
    km s$^{-1}$}), a medium-broad emission profile \mbox{(FWHM
    $\sim$5--10 km s$^{-1}$)}, and narrow self-absorption lines
  \mbox{(FWHM $\sim$2--3 km s$^{-1}$)}; see the H$_2$O $2_{02}$--$1_{11}$
  lines in Fig.~\ref{3:fig:AllSpectraOntop}. The same components are
  also seen in the CO line profiles, albeit less prominently than for 
  H$_2$O.  The broad component dominates the
  \mbox{$^{12}$CO 10--9} lines of IRAS~4A and 4B and is also apparent
  in the deep \mbox{$^{12}$CO 6--5} spectrum of IRAS~2A
  (Fig.~\ref{3:fig:ratios}). The medium component is best seen in the
  \mbox{$^{13}$CO 10--9} profiles of IRAS~4A and 4B and as the red
  wing of the $^{12}$CO 10--9 profile for IRAS~2A. A blow-up of the
  very high $S/N$ spectrum of C$^{18}$O 5--4 for IRAS~4A 
   (insert in Fig.~\ref{3:fig:AllSpectraOntop}) also reveals a weak
  C$^{18}$O medium-broad profile.
The narrow component is clearly observed in
C$^{18}$O emission and $^{12}$CO \mbox{low-$J$} self-absorption.
\citet{Kristensen10}  interpret the broad component as
shocked gas along the outflow cavity walls, the medium component
as smaller-scale shocks created by the outflow in the inner ($<$1000
AU) dense envelope, and the narrow component as the quiescent envelope,
respectively.



\section{Analysis and discussion}

\subsection{Broad and medium components: shocked gas}

To quantify the physical properties of the broad outflow component,
line ratios are determined for the wings of the line profiles. Figure
\ref{3:fig:ratios} shows the \mbox{\element[]{CO}
  6--5}/\mbox{\element[]{CO} 10--9} ratio as a function of velocity.
The APEX-CHAMP$^+$ \mbox{\element[]{CO} 6--5} maps of IRAS~4A and 4B
from Y{\i}ld{\i}z et al. (in prep.) and IRAS~2A from
\citet{vanKempen09champ2} are resampled
to a 20$\arcsec$ beam so that both lines refer to the same beam. The
ratios are compared with model non-LTE excitation line intensities
calculated using the \verb1RADEX1 code \citep{2007A&A...468..627V}
(Fig. \ref{3:fig:RadexModel}, Online Appendix A).  The density within a
20$\arcsec$ diameter is taken to be $\ge$10$^{5}$ cm$^{-3}$ based on the
modeling results of \citet[][see also Sect. 3.3 and Appendix A]{Jorgensen02}.
The detection of medium-broad \mbox{\element[]{CS} 10--9} emission by
\citet{Jorgensen05} toward IRAS~4A and 4B indicates densities of the order of
a few $10^6$ cm$^{-3}$. For the range of densities indicated in
Fig.~\ref{3:fig:RadexModel}, the line ratios
imply high temperatures: IRAS~2A, \mbox{$T_{\mathrm{kin}}=$70--130 K};
IRAS~4A, \mbox{$T_{\mathrm{kin}}=$90--120 K}; and IRAS~4B,
\mbox{$T_{\mathrm{kin}}=$140--180 K}.

The optical depth of the \element[][12]{CO} emission is constrained by
the \mbox{\element[][12]{CO} 10--9}/\mbox{\element[][13]{CO} 10--9}
ratios.  For IRAS~4B, the optical depth of the \element[][12]{CO} line
wings is found to drop with velocity, ranging from
$\tau_{\mathrm{wing}}\sim$12 near the center to $\sim$0.4 at the
highest velocities where \element[][13]{CO} is detected. This
justifies the assumption that the broad
\mbox{\element[][12]{CO}~10--9} lines are optically thin.
Total CO column densities in the broad component for these conditions
are 4 and \mbox{1 $\times 10^{16}$ cm$^{-2}$} for IRAS~4A and 4B,
respectively.  For IRAS~2A, the broad column density is
calculated from the \mbox{CO~6--5} spectrum as \mbox{6 $\times
  10^{15}$ cm$^{-2}$}. Using \mbox{\element[]{CO}/H$_{2}$ = $10^{-4}$}
gives the H$_{2}$ column densities listed in Table
\ref{tbl:columndensities}.

\placeFigureRatios

The medium component attributed to small-scale shocks in the
inner envelope can be probed directly by the
\mbox{\element[][13]{CO} 10--9} data for IRAS~4A and 4B. For
IRAS~2A, the Gaussian fit to the red wing of the $^{12}$CO 10--9 is used.
By assuming a similar
range of temperatures and densities as for the broad component,
beam averaged \mbox{\element[][12]{CO}} column densities of 2, 6,
  and \mbox{2 $\times10^{16}$~cm$^{-2}$} are found for IRAS~2A, 4A,
  and 4B respectively, if the lines are optically thin and using
\mbox{$^{12}$C/$^{13}$C = 65}. The very weak medium component found in
the C$^{18}$O 5--4 profile for IRAS~4A agrees with this value if the
emission arises from a compact (few $\arcsec$) source. Assuming
CO/H$_2$=10$^{-4}$ leads to the numbers in Table
\ref{tbl:columndensities}. The overall uncertainty in all
  column densities is a factor of 2 due to
  the range of physical conditions used to derive them and
uncertainties in the adopted CO/H$_2$ ratio and calibration. The total amount 
of shocked gas is $<$1\% of the total gas column density in the beam for each source
\citep{Jorgensen02}.

\subsection{Narrow component: bulk warm envelope}
\placeTableColumnDensities
\placeFigureRatran

The narrow width of the C$^{18}$O emission clearly indicates an origin
in the quiescent envelope.  Na\"ively, one would associate emission
coming from a level with $E_\mathrm{u}/k_\mathrm{B}$= 237~K (9--8)
with the warm gas in the innermost part of the envelope. To test this
hypothesis, a series of envelope models was run with varying CO
abundance profiles.  The models were constructed assuming a power-law
density structure and then calculating the temperature structure by
fitting both the far-infrared spectral energy distribution and the
submillimeter spatial extent \citep{Jorgensen02}.  Figure
\ref{3:fig:RatranModel} compares the fractional line intensities for
the C$^{18}$O and $^{13}$CO transitions in a spherical envelope model
for IRAS~2A as a function of temperature.  In these models, the
abundance in the outer envelope was kept high,
$X_{0}$=$2.7\times 10^{-4}$ with respect to H$_2$ (all
available gas-phase carbon in CO), decreasing by a factor of 1000 at
temperatures higher than a specific temperature, $T_0$ (a so-called
`anti-jump' model \citep[see][for nomenclature]{Schoeier04}. These
models thereby give an estimate of the fraction of the line emission
for a given transition (in the respective telescope beams) which has
its origin at temperatures lower than $T_0$.

For C$^{18}$O, 90\% of the emission in the transitions up to and
including the 5--4 HIFI transition has its origin at
temperatures lower than 25--30~K, meaning that these transitions are
predominantly sensitive to the outer parts of the protostellar
envelope. The 9--8 transition is more sensitive to the warm parts
of the envelope, but still 50\% of the line flux appears to come from
the outer envelope with temperatures less than 50~K. The $^{13}$CO
transitions become rapidly optically thick in the outer envelopes:
even for the 9--8 transition, 90\% of the line flux can be
associated with the envelope material with temperatures lower than
40~K.

The C$^{18}$O 9--8 line 
is clearly a much more sensitive probe of a CO ice
evaporation zone than any other observed CO line. \citet{Jorgensen05freeze}
showed that the low-$J$ C$^{18}$O lines require a drop in the
abundance at densities higher than $7\times 10^4$ cm$^{-3}$ due to
freeze-out. However, they did not have strong proof for CO evaporation
in the inner part from that dataset. Using the temperature and density
structure for IRAS~2A as described above, we computed the
C$^{18}$O 
line intensities in the respective telescope beams following the
method by \citet{Jorgensen05freeze}.  In this `anti-jump' model,
 the outer C$^{18}$O abundance is kept fixed at $X_0=5.0\times 10^{-7}$, 
whereas the inner abundance $X_{\rm D}$ and
the freeze-out density $n_{\rm de}$ are free parameters.
A $\chi^2$ fit to only the \mbox{C$^{18}$O 1--0}, 2--1 and 3--2 lines 
gives best-fit values of $X_{\rm D}=3\times 10^{-8}$
and $n_{\rm de}=7\times 10^4$ cm$^{-3}$, consistent with those
of \citet{Jorgensen05freeze}. 
However, this model underproduces the higher-$J$ lines by a factor of 3--4
  (Fig.~\ref{overplot1} in Appendix B).

To solve this underproduction, the inner
abundance has to be increased in a so-called `drop-abundance' profile.
The fit parameters are now the inner abundance $X_{\rm in}$ and the
evaporation temperature $T_{\rm ev}$, keeping $X_{\rm D}$ and $n_{\rm de}$ fixed at
the above values.
Figure \ref{Chi2plots} in Appendix B shows the $\chi^2$ plots to the
C$^{18}$O 6--5 and 9--8 lines.  The evaporation temperature is not
well constrained, but low temperatures of $T_{\rm ev}\approx$ 25~K are
favored because they produce more \mbox{C$^{18}$O 5--4} emission. 
The best-fit \mbox{$X_{\rm
    in}= 1.5\times 10^{-7}$} indicates a jump of a factor of 5
compared with $X_{\rm D}$. Alternatively, $T_{\rm ev}$ can be kept
fixed at 25 K and both $X_{\rm in}$ and $X_{\rm D}$ can be varied by
fitting all five lines simultaneously. In this case, the same best-fit
value for $X_{\rm in}$ is found but only an upper limit on $X_{\rm D}$
of $\sim 4\times 10^{-8}$.  Thus, for this physical model,
$X_{\rm in}>X_{\rm D}$,
implying that a jump in the abundance is needed for IRAS~2A.

\section{Conclusions}

Spectrally resolved \textit{Herschel}-HIFI observations of high-$J$ CO
lines up to \mbox{\element[][12]{CO} 10--9} and
\mbox{C\element[][18]{O} 9--8} have been performed toward three
low-mass young stellar objects for the first time. These data provide
strong constraints on the density and temperature in the various
physical components, such as the quiescent envelope, extended
outflowing gas, and small-scale shocks in the inner envelope. The
derived column densities and temperatures are important for
comparison with water and other molecules such as O$_2$, for which
HIFI observations are planned. Furthermore, it is shown conclusively
that in order to reproduce higher-$J$ C$^{18}$O lines within
  the context of the adopted physical model, a jump in the CO
abundance due to evaporation is required in the inner envelope,
something that was inferred, but not measured, from ground-based
observations.  Combination with even higher-$J$ \element[]{CO} lines
to be obtained with \textit{Herschel-}PACS in the frame of the WISH
key program will allow further quantification of the different
physical processes invoked to explain the origin of the high-$J$
emission.


\bibliographystyle{aa} 
\bibliography{bibdata}

\begin{acknowledgements}
The authors are grateful to many funding agencies and the HIFI-ICC staff who has been contributing for the construction of \textit{Herschel} and HIFI for many years. HIFI has been designed and built by a consortium of institutes and university departments from across Europe, Canada and the United States under the leadership of SRON Netherlands Institute for Space
Research, Groningen, The Netherlands and with major contributions from Germany, France and the US.
Consortium members are: Canada: CSA, U.Waterloo; France: CESR, LAB, LERMA, IRAM; Germany:
KOSMA, MPIfR, MPS; Ireland, NUI Maynooth; Italy: ASI, IFSI-INAF, Osservatorio Astrofisico di Arcetri-
INAF; Netherlands: SRON, TUD; Poland: CAMK, CBK; Spain: Observatorio Astron{\'o}mico Nacional (IGN),
Centro de Astrobiolog{\'i}a (CSIC-INTA). Sweden: Chalmers University of Technology - MC2, RSS $\&$ GARD;
Onsala Space Observatory; Swedish National Space Board, Stockholm University - Stockholm Observatory;
Switzerland: ETH Zurich, FHNW; USA: Caltech, JPL, NHSC.
\end{acknowledgements}

\Online
\appendix
\section{Radex model}
\begin{figure}
\begin{center}
\includegraphics[scale=0.4]{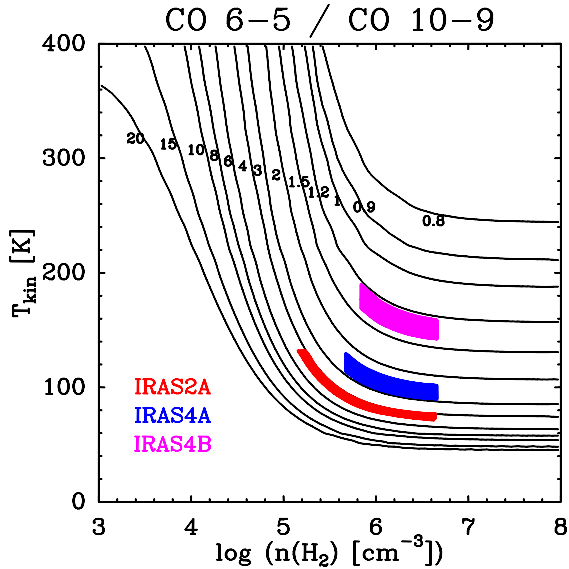}
\end{center}
\caption{Model line ratios of CO 6--5/10--9 for a slab model with a
  range of temperatures and densities. The adopted CO column density is
  10$^{17}$ cm$^{-2}$ with a line width of 10 km s$^{-1}$, comparable 
to the inferred
values. For these parameters
  the lines involved are optically thin. The colored lines give the 
range of densities within the 20$''$ beam for the three sources based on the
models of \citet{Jorgensen02}.}
\label{3:fig:RadexModel}
\end{figure}

Figure \ref{3:fig:RadexModel} shows the CO 6--5/10--9 line ratios for a slab model with a range of temperatures and densities.

\section{Abundance profiles for IRAS~2A}
Among the three sources, IRAS~2A has been selected for detailed CO
abundance profile modeling because more data are available on this
source, and because its physical and chemical structure has been well
characterized through the high angular resolution submillimeter single
dish and interferometric observations of
\citet{Jorgensen02,Jorgensen05i2}.  The physical parameters are taken
from the continuum modeling results of \citet{Jorgensen02}. In that
paper, the 1D dust radiative transfer code \verb1DUSTY1
\citep{ivezic_elitzur97} was used assuming a power law to describe the
density gradient. The dust temperature as function of radius was
calculated self-consistently through radiative transfer given a
central source luminosity. Best-fit model parameters were obtained by
comparison with the spectral energy distribution and the submillimeter
continuum spatial extent. The resulting envelope structure parameters
are used as input to the \verb1Ratran1 radiative transfer modeling
code \citep{Hogerheijde00} to model the CO line intensities for a
given CO abundance structure through the envelope. The model
  extends to 11000 AU from the protostar, where the density has dropped
  to $2\times 10^4$ cm$^{-3}$. The CO-H$_{2}$ collisional rate
coefficients of \citet{Yang10} have been adopted.

\begin{figure}[htb]
   \centering
    \includegraphics[scale=0.35, angle=270]{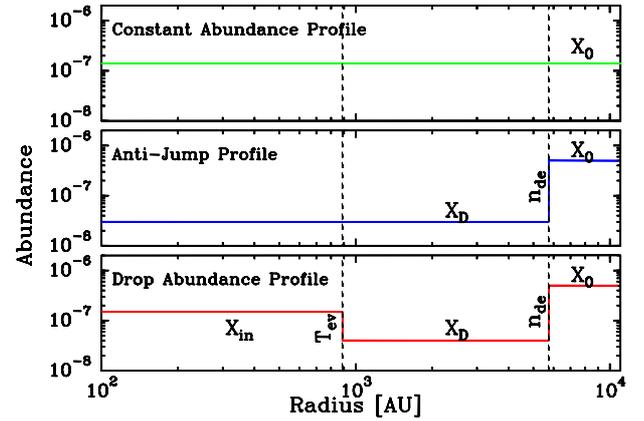}
    \caption{\small Examples of constant, anti-jump, and drop abundance profiles 
    for IRAS~2A for $T_{\rm ev}$=25 K and $n_{\rm de}$=7$\times$10$^4$ cm$^{-3}$}.
    \label{model1}
\end{figure}

The C$^{18}$O lines are used to determine the CO
  abundance structure because the lines of this isotopologue are
  largely optically thin and because they have well-defined Gaussian
  line shapes originating from the quiescent envelope without
  strong contaminations from outflows. Three types of abundance profiles are
  examined, namely `constant', `anti-jump' and `drop' abundance
  profiles. Illustrative models are shown in
  Fig. \ref{model1} and the results from these models are summarized in Table \ref{model1}.
  
\begin{table}
\caption{Summary of CO abundance profiles for IRAS~2A.}
\tiny
\begin{center}
\begin{tabular}{l c c c c c }
\hline \hline
Profile      & $X_{\rm in}$ & $T_{\rm ev}$ & $X_{\rm D}$ & $n_{\rm de}$ & $X_{0}$ \\
                &                      & (K)          &                     & (cm$^{-3}$)   &              \\
\hline
Constant    & -  & - & - & - &  1.4$\times$10$^{-7}$ \\
Anti-jump  & -  & - & 3$\times$ 10$^{-8}$ & 7$\times$10$^4$ &  5$\times$10$^{-7}$ \\
Drop          &  1.5$\times$10$^{-7}$  & 25 & $\sim$ 4$\times$10$^{-8}$ & 7$\times$10$^4$ & 5$\times$10$^{-7}$   \\
\hline
\end{tabular}
\end{center}
\label{tbl:abundancetable}
\end{table}

\subsection{Constant abundance model}
The simplest approach is to adopt a constant abundance across the entire envelope. However, with this approach, and within the framework of the adopted source model, it is not possible to simultaneously reproduce all line intensities. This was already shown by \citet{Jorgensen05freeze}. For lower abundances it is possible to reproduce the lower-$J$ lines, while higher abundances are required for higher-$J$ lines. In Fig. \ref{overplot1} the C$^{18}$O spectra of a constant-abundance profile are shown for an abundance of $X_0$=1.4$\times$10$^{-7}$, together with the observed spectra of IRAS 2A. Based on these results, the constant-abundance profile is ruled out for all three sources.

\subsection{Anti-jump abundance models}
The anti-jump model is commonly adopted in
  models of pre-stellar cores without a central heating source
  \citep[e.g.,][]{Bergin02,Tafalla04}. Following \citet{Jorgensen05freeze}, an anti-jump abundance
  profile was employed by varying the desorption density, $n_{\rm
    de}$, and inner abundance $X_{\rm in}$=$X_{\rm D}$ in order to
  find a fit to our observed lines. Here, the outer abundance
  $X_{0}$ was kept high at 5.0$\times$10$^{-7}$ corresponding to a
  $^{12}$CO abundance of 2.4$\times$10$^{-4}$ for
  $^{16}$O/$^{18}$O=550 as was found appropriate for the 
  case of IRAS~2A by \citet{Jorgensen05freeze}. This value is consistent 
  with the CO/H$_{2}$ abundance ratio determined by 
  \citet{Lacy94} for dense gas without CO freeze-out. 

\begin{figure}
   \centering
    \includegraphics[scale=0.5]{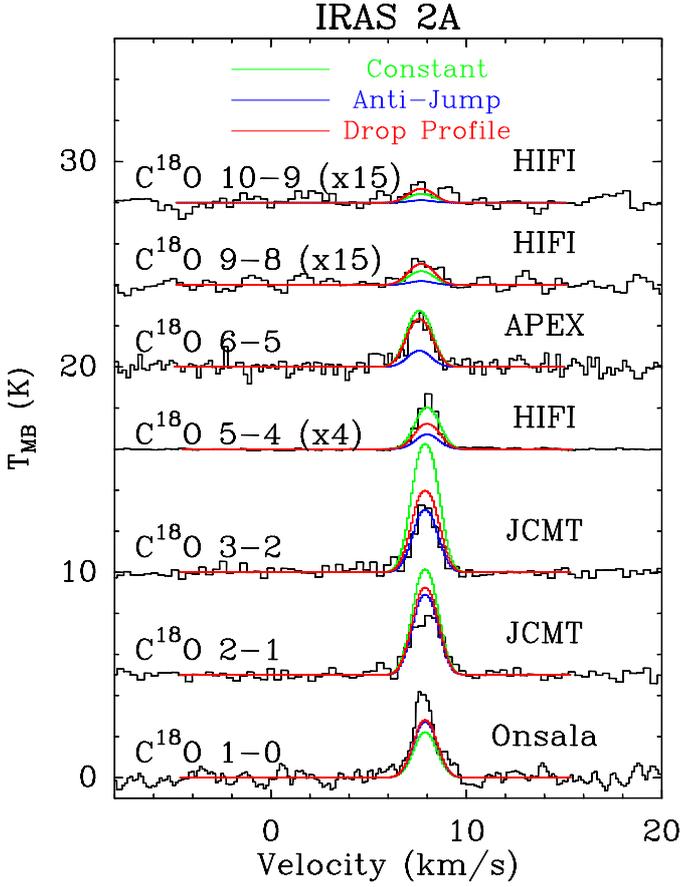}
    \caption{\small  Best fit constant (green), anti-jump (blue) and drop abundance (red)
      Ratran models overplotted on the observed spectra. All
        spectra refer to single pointing observations.  The
        calibration uncertainty for each spectrum is around 20--30$\%$
        and is taken into account in the $\chi^2$ fit. See Table~\ref{tbl:abundancetable} for parameters.}
    \label{overplot1}
\end{figure}

The best fit to the three lowest C$^{18}$O lines (1--0, 2--1 and 3--2)
is consistent with that found by \citet{Jorgensen05freeze},
corresponding to $n_{\rm de}$=7$\times$10$^{4}$ cm$^{-3}$ and $X_{\rm
  D}$=3$\times$10$^{-8}$ (CO abundance of $1.7\times
10^{-5}$). In the $\chi^{2}$ fits, the calibration uncertainty
  of each line (ranging from 20 to 30$\%$) is taken into account.
These modeled spectra are overplotted on the observed spectra in
Fig. \ref{overplot1} as the blue lines, and show that the anti-jump
profile fits well the lower-$J$ lines but very much underproduces the
higher-$J$ lines.

\begin{figure}
   \centering
    \includegraphics[scale=0.24]{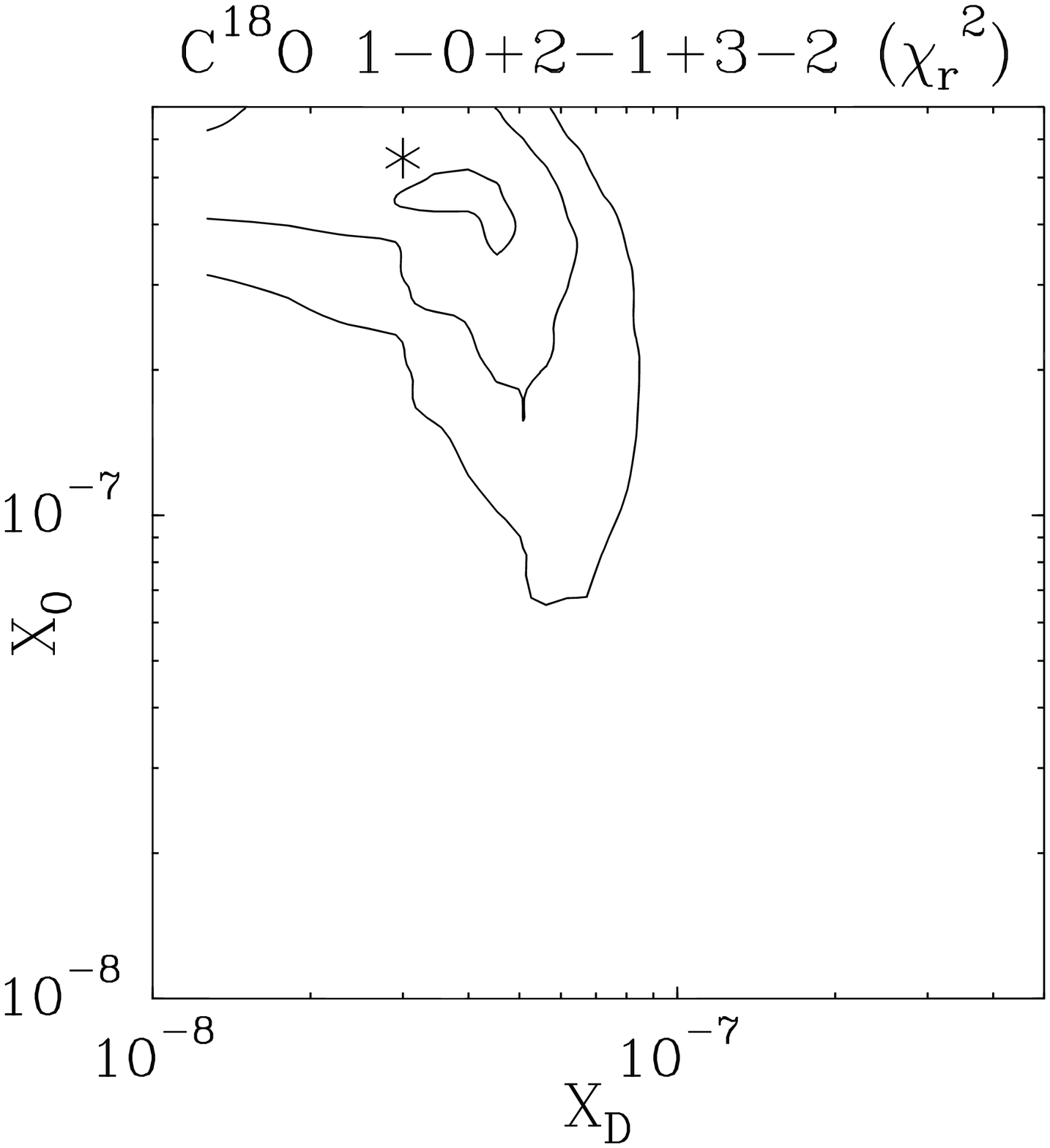}
    \includegraphics[scale=0.24]{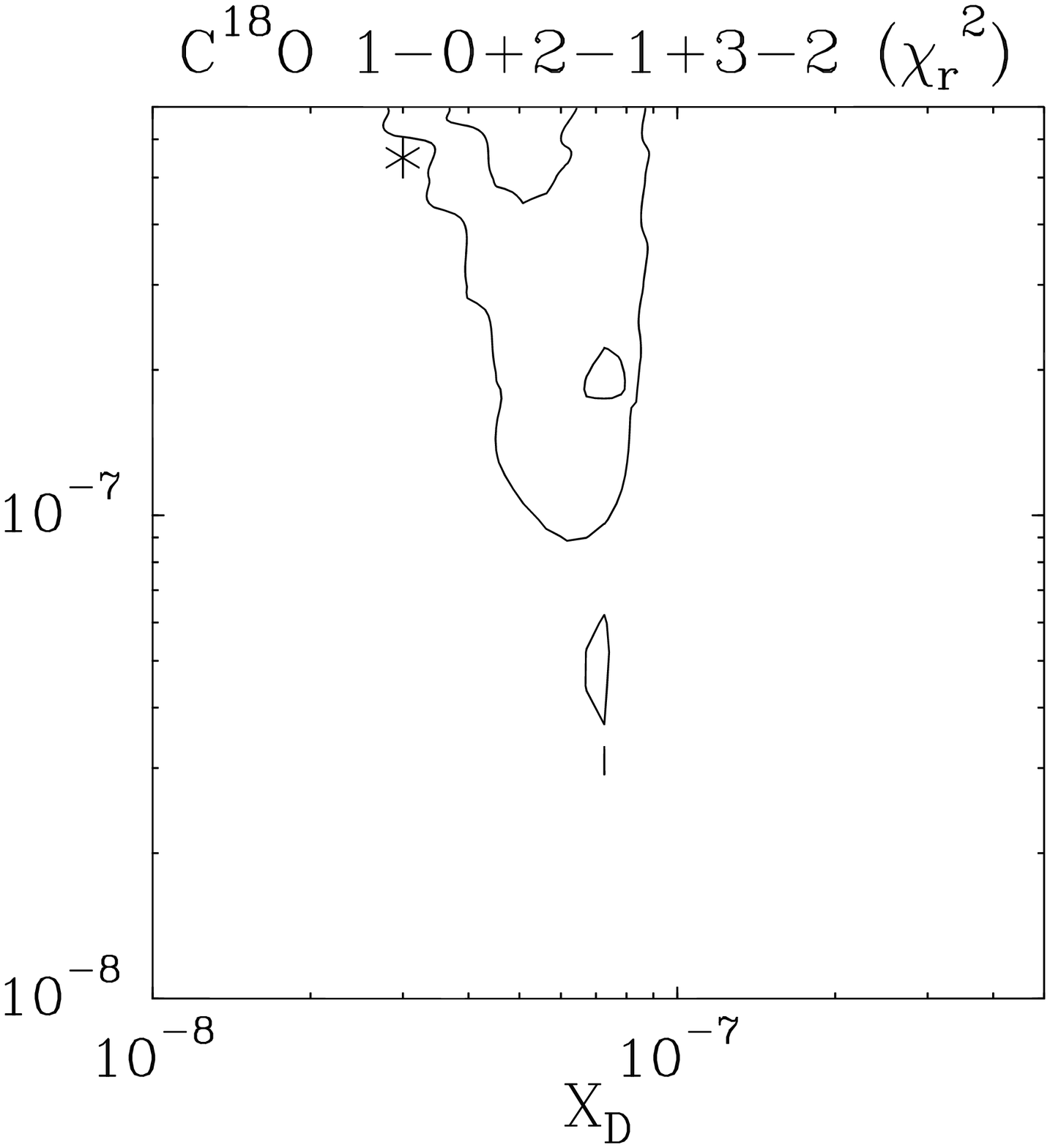}
    \caption{\small The $\chi^{2}$ plots for the anti-jump profiles
      where $X_0$ and $X_{\rm D}$ values are varied. {\it Right:} for
      $n_{\rm de}$=7$\times$10$^{4}$ and {\it left:} for $n_{\rm
        de}$=3.4$\times$10$^{4}$ cm$^{-3}$. The asterisk indicates the
      value for \citet{Jorgensen05freeze} used here. Contours are plotted at
 the 2$\sigma$, 3$\sigma$, and 4$\sigma$ confidence levels (\textit{left}) and 3$\sigma$ and 4$\sigma$ confidence levels (\textit{right}).}
    \label{referee1}
\end{figure}

\begin{figure}
   \centering
    \includegraphics[scale=0.245]{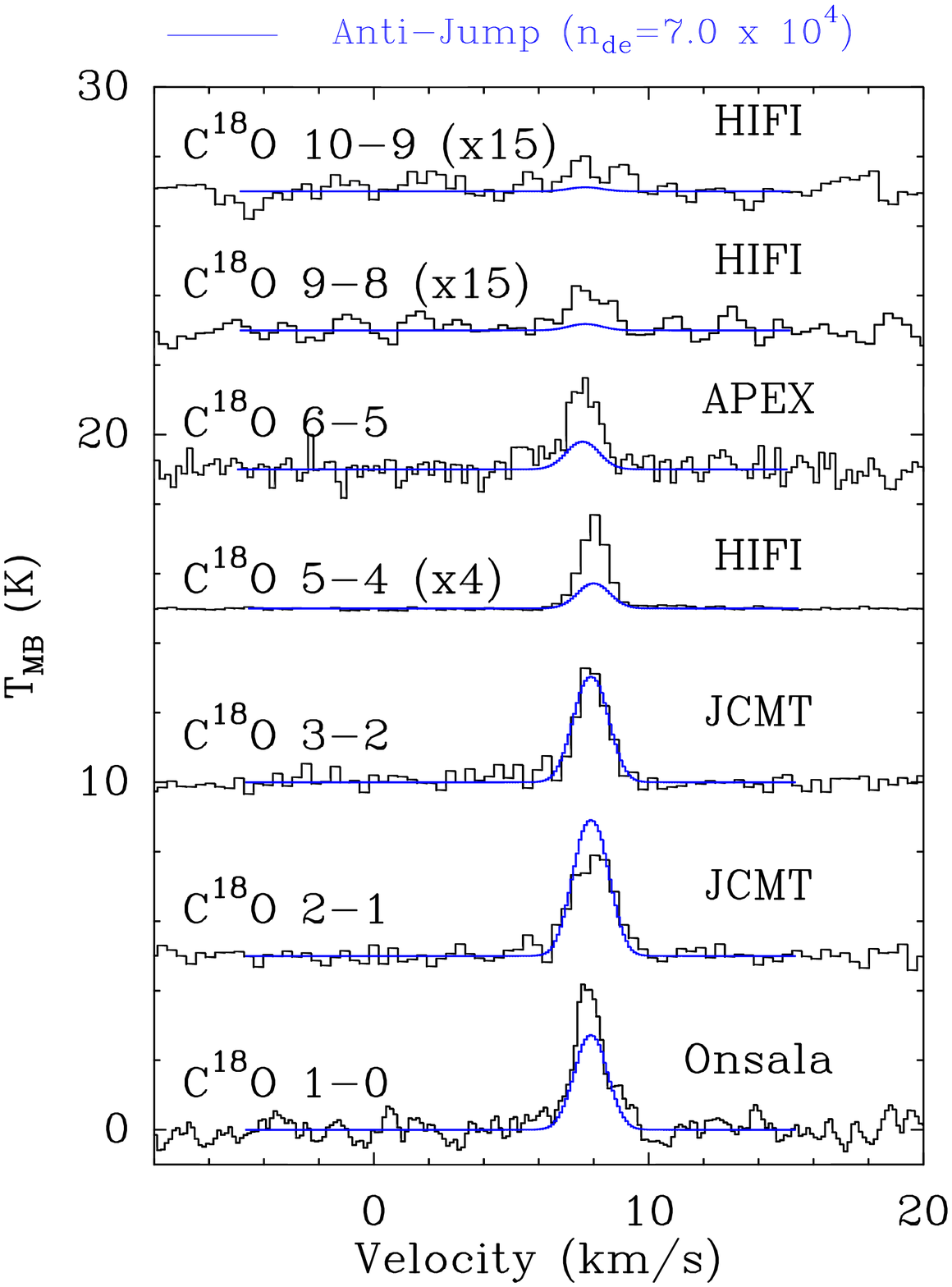}
    \includegraphics[scale=0.245]{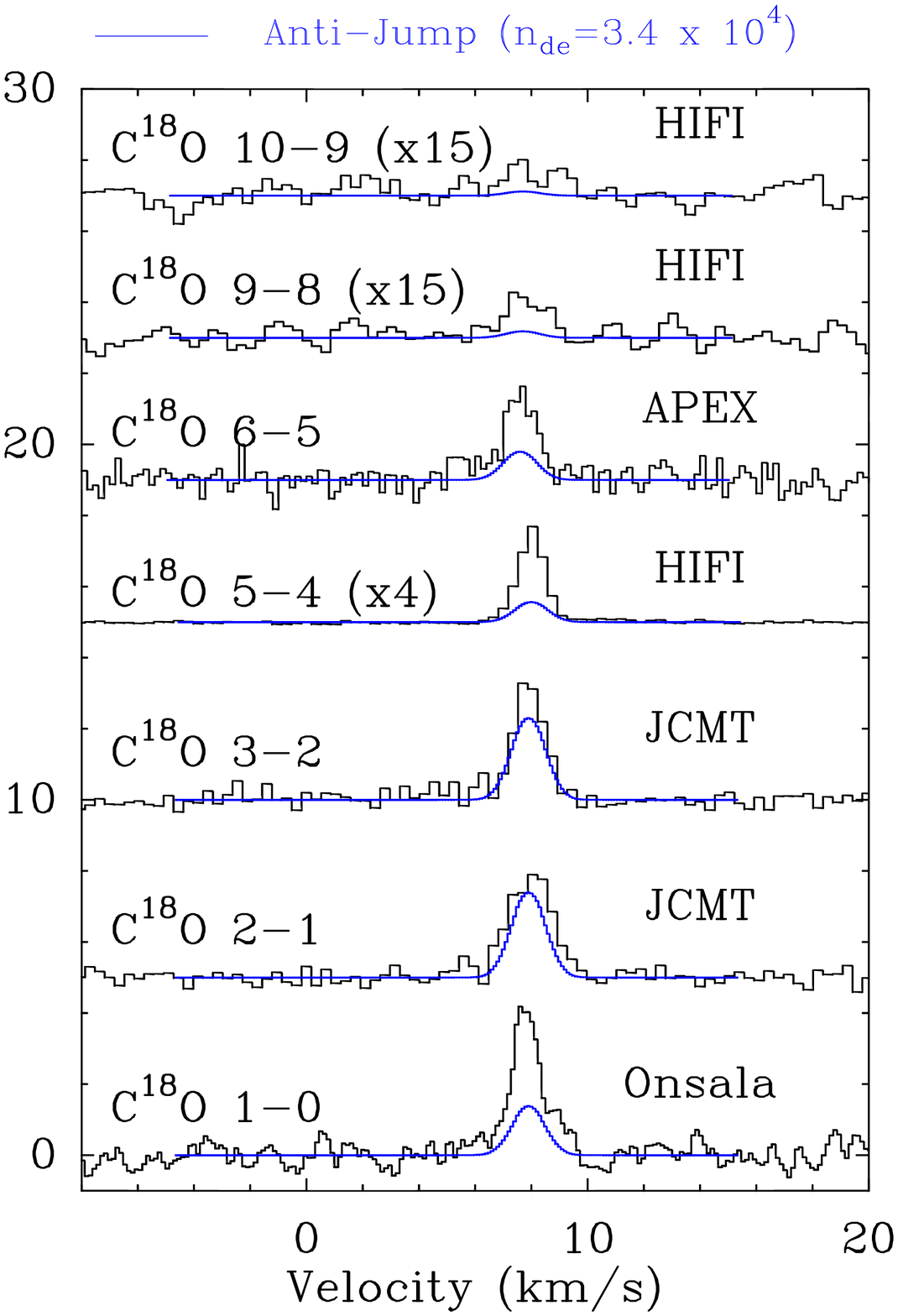}
    \caption{\small The IRAS~2A spectra for the $X_0$ and $X_{\rm D}$
      parameters corresponding to the values in \citet{Jorgensen05freeze} for different $n_{\rm de}$ values of 3.4$\times$10$^{4}$ and 7$\times$10$^{4}$ cm$^{-3}$.}
    \label{referee2}
\end{figure}

The value of $X_0$ was verified a posteriori by keeping $n_{\rm de}$ at two different values of 3.4$\times$10$^{4}$ and 7$\times$10$^{4}$ cm$^{-3}$. This is illustrated in Fig. \ref{referee1} where the $\chi^2$ contours show that for both values of $n_{\rm de}$, the best-fit value of $X_0$ is $\sim$5$\times$10$^{-7}$, the value also found in \citet{Jorgensen05freeze}. The $\chi^2$ contours have been calculated from the lower-$J$ lines only, as these are paramount in constraining the value of $X_0$. Different $\chi^2$ plots were made, where it was clear that higher-$J$ lines only constrain $X_{\rm D}$, as expected. The effect of $n_{\rm de}$ is illustrated in Fig. \ref{referee2} for the two values given above.

\begin{figure}
   \centering
    \includegraphics[scale=0.24]{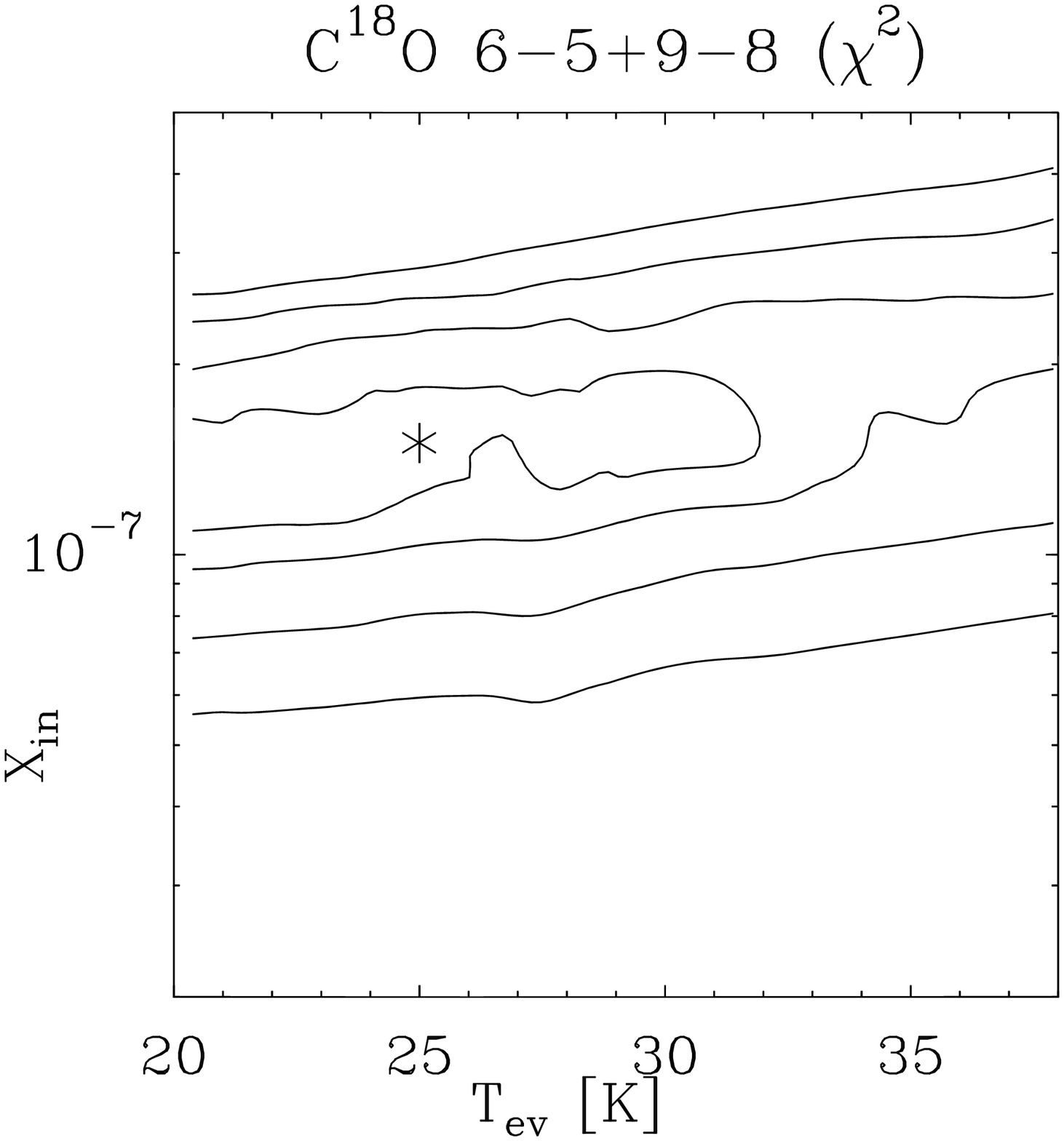}
    \includegraphics[scale=0.24]{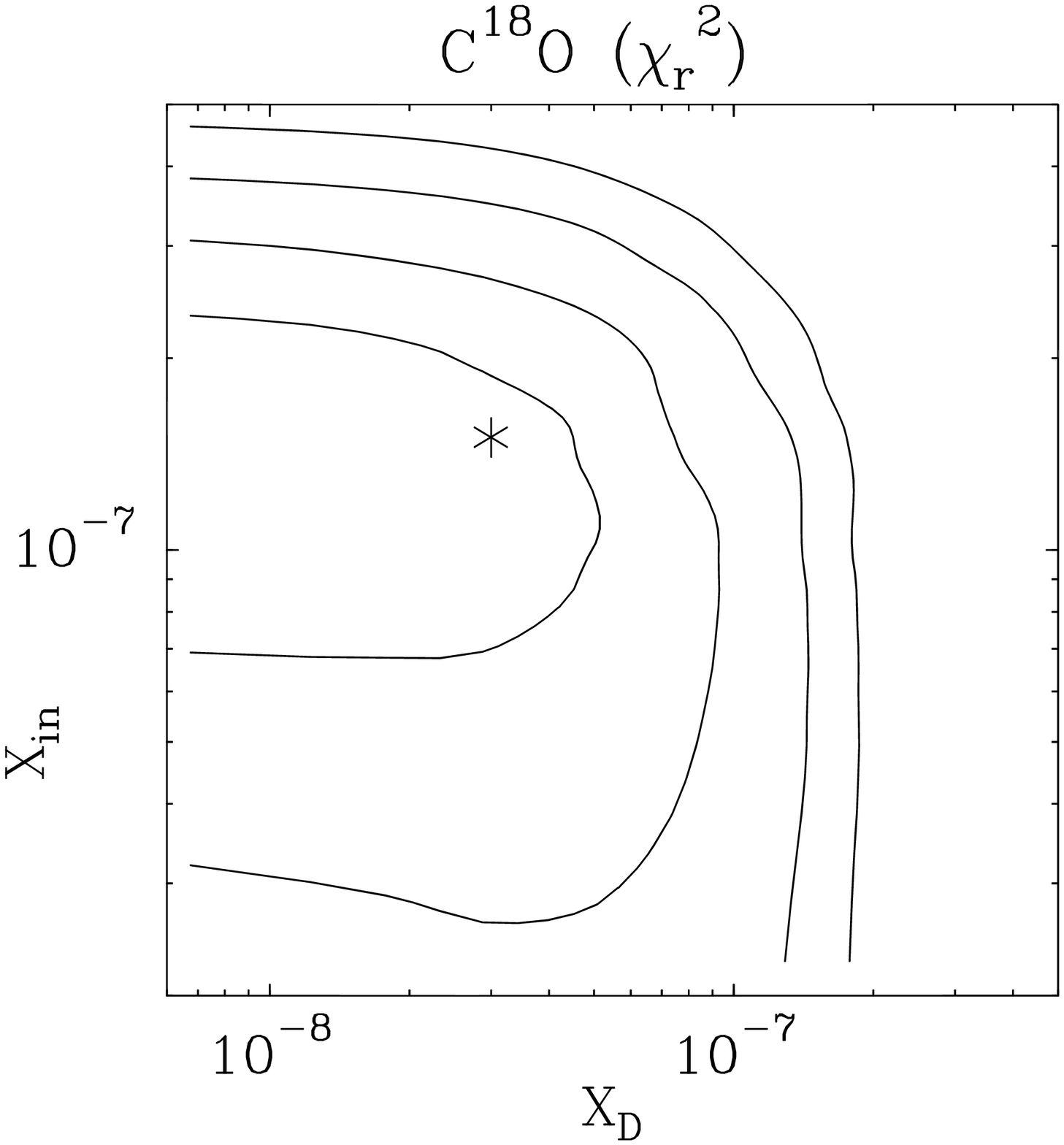}
    \caption{\small Reduced $\chi^2$ plots
and best-fit parameters (indicated with *)
for the anti-jump model 
fit to the lines of  C$^{18}$O 1--0, 2--1, 3--2, 6--5 and 9--8 (\textit{right})
and for the drop abundance model fit to the higher-$J$ lines 
of  C$^{18}$O 6--5 and 9--8 (\textit{left}). Contours are plotted at
 the 1$\sigma$, 2$\sigma$, 3$\sigma$, and 4$\sigma$ confidence levels.}
    \label{Chi2plots}
\end{figure}

\subsection{Drop-abundance profile}

In order to fit the higher-$J$ lines, it is necessary to employ a
  drop-abundance structure in which the inner abundance $X_{\rm in}$
  increases above the ice evaporation temperature $T_{\rm ev}$
  \citep{Jorgensen05freeze}. The abundances $X_{\rm D}$ and $X_0$ for
  $T<T_{\rm ev}$ are kept the same as in the anti-jump model, but
  $X_{\rm in}$ is not necessarily the same as $X_{0}$. In order to
  find the best-fit parameters for the higher-$J$ lines, the inner
  abundance $X_{\rm in}$ and the evaporation temperature $T_{\rm ev}$
  were varied. The $\chi ^{2}$ plots (Fig. \ref{Chi2plots}, left
  panel) show best-fit values for an inner abundance of \mbox{$X_{\rm
      in}$ = 1.5$\times$10$^{-7}$} and an evaporation temperature of
  25 K (consistent with the laboratory values), although the latter
  value is not strongly constrained. These parameters fit well the
  higher-$J$ C$^{18}$O 6-5 and 9-8 lines (Fig. \ref{overplot1}). The
  \mbox{C$^{18}$O 5--4} line is underproduced in all models, likely
  because the larger HIFI beam picks up extended emission from 
    additional dense material to the northeast of the source seen in
  BIMA \mbox{C$^{18}$O 1--0} \citep{Volgenau06} map.

Because the results do not depend strongly on $T_{\rm ev}$, an
  alternative approach is to keep the evaporation temperature fixed at
  25 K and vary both $X_{\rm in}$ and $X_{\rm D}$ by fitting both low-
  and high-$J$ lines simultaneously. In this case, only an upper limit
  on $X_{\rm D}$ of $\sim 4\times 10^{-8}$ is found
  (Fig. \ref{Chi2plots}, right panel), whereas the inferred value of
  $X_{\rm in}$ is the same.  This figure conclusively illustrates that
  $X_{\rm in}>X_{\rm D}$, i.e., that a jump in the abundance due to
  evaporation is needed.

 The above conclusion is robust within the context of the
  adopted physical model. Alternatively, one could investigate
  different physical models such as those used by \citet{Chiang08},
  which have a density enhancement in the inner envelope due to a
  magnetic shock wall. This density increase could partly mitigate the
  need for the abundance enhancement although it is unlikely that the
  density jump is large enough to fully compensate. Such models are
  outside the scope of this paper. An observational test of our model
  would be to image the N$_2$H$^+$ 1--0 line at high angular
  resolution: its emission should drop in the inner $\sim$900 AU
  ($\sim$4$\arcsec$) where N$_2$H$^+$ would be destroyed by the enhanced
  gas-phase CO.

\end{document}